\documentclass[prd,aps,showpacs,preprintnumbers,amssymb]{revtex4}
\usepackage{axodraw}
\usepackage{color}
\usepackage{epsf}
\usepackage{graphicx}

\def\e3p{$\eta \rightarrow 3 \pi$}

\begin{document}
\title{%
\hfill{\normalsize\vbox{%
\hbox{}
 }}\\
{An analytical treatment of the neutrino masses and mixings}}
\author{Renata Jora
$^{\it \bf a}$~\footnote[1]{Email:
 rjora@theory.nipne.ro}}

\author{Joseph Schechter
 $^{\it \bf b}$~\footnote[2]{Email:
 schechte@phy.syr.edu}}

\author{M. Naeem Shahid
$^{\it \bf c}$~\footnote[3]{Email:
   mnshahid@phy.syr.edu }}

\affiliation{$^{\bf \it a}$ National Institute of Physics and Nuclear Engineering PO Box MG-6, Bucharest-Magurele, Romania}
\affiliation{$^ {\bf \it b}$ Department of Physics,
 Syracuse University, Syracuse, NY 13244-1130, USA}
\affiliation{$^ {\bf \it c}$ National Centre for Physics,
 Quaid-i-Azam University Campus, 45320 Islamabad, Pakistan}

\date{\today}

\begin{abstract}
We obtain analytical formulas which connect the neutrino masses and the leptonic mixing matrix with the entries in the
 mass matrix for the approximation in which the charged lepton mixing matrix is the unit matrix. We also extract the CP violation phase and determine the conditions in which this is present.
\end{abstract}
\maketitle

\section{Introduction}

The problem of neutrino masses and mixings has received  a surge of interest as a consequence of a remarkable number of experiments \cite{K}-\cite{MINOS} involving neutrino oscillations.
In a great number of models proposed in connection to it it is safe to assume that at least in first orders the mixing matrix in the charged lepton sector is the unit matrix.

Consider that part of the Lagrangian which contains the lepton fields and their interaction with the $SU(2)_L\times U(1)$ gauge bosons:
\begin{eqnarray}
{\cal L}=-{\bar L}_a\gamma_{\mu}{\cal D}_{\mu}L_a-{\bar e}_{Ra}\gamma_{\mu}{\cal D}_{\mu}e_{Ra}
\label{lagr444}
\end{eqnarray}
Here,
\begin{eqnarray}
L_a=
\left[
\begin{array}{c}
\rho_a\\
e_{La}
\end{array}
\right],
\label{l665}
\end{eqnarray}

and $\rho_a$ is a two component neutrino field.

The charged leptonic weak interaction can be read from the term:
\begin{eqnarray}
{\cal L}=\frac{i g}{\sqrt{2}}W_{\mu}^{-}{\hat {\bar{e}}}_L\gamma_{\mu}K\hat{\rho}+h.c.+...,
\label{lept7776}
\end{eqnarray}

where g is the weak coupling constant, $W_{\mu}^-$ is the charged vector intermediate boson and K is the leptonic mixing matrix.

The hatted fields correspond to the mass eigenstates and can be written in terms of the gauge eigenstates as:
\begin{eqnarray}
\rho=U\hat{\rho}
\hspace{1cm}
e_L=W{\hat{e}}_L.
\label{gg55}
\end{eqnarray}

The leptonic mixing matrix K is then given by:
\begin{eqnarray}
K=W^{\dagger}U.
\label{lept66}
\end{eqnarray}

If W is the unit matrix then the leptonic mixing matrix is given simply by U.

One then can consider arbitrary mass terms in the neutrino sector and diagonalize them. At first glance this seems a tractable problem considering that it involves only $3\times3$
 arbitrary complex matrices. In practice this diagonalization, upon obtaining $MM^{\dagger}$, involves 9 parameters and it is often made numerically. In the case when various symmetries and
 perturbations of these are applied in the neutrino sector \cite{Wolfenstein}-\cite{Jora4} one may lose track of the real dependence between the input mass parameters and the output eigenvalues and eigenvectors.

 In the present work we obtain analytical formulas relating the various masses  and angles in the neutrino sector in terms of the input mass parameters. We also extract
 the possible CP violating phase and determine the necessary and sufficient conditions for its appearance. Note that overall we made the underlying assumption that the charged leptons mixing matrix is the unit matrix.
\section{Determination of parameters}
We consider the neutrino mass matrix an arbitrary $3\times3$ matrix  $M_{\nu}$ from which we obtain the hermitian form $M=M_{\nu}M^{\dagger}_{\nu}$:

\begin{eqnarray}
M=
\left[
\begin{array}{ccc}
a&x&y\\
x^*&b&z\\
y^*&z^*&c
\end{array}
\right]
\label{mass5454}
\end{eqnarray}

Here the entries a, b, c are real and,
\begin{eqnarray}
&&x=|x|e^{i\beta_1}
\nonumber\\
&&y=|y|e^{i\beta_2}
\nonumber\\
&&z=|z|e^{i\beta_3}.
\label{res44332}
\end{eqnarray}

The matrix M is diagonalized by an arbitrary unitary matrix U. If U is identified with the leptonic mixing matrix than U is equivalent to \cite{Schechter1}, \cite{Schechter2}

\begin{eqnarray}
U\rightarrow P^*UQ,
\label{res4443}
\end{eqnarray}

where P and Q are diagonal matrices of phases.

We define the phase matrix P by,

\begin{eqnarray}
P=
\left[
\begin{array}{ccc}
e^{i\gamma_1}&0&0\\
0&e^{i\gamma_2}&0\\
0&0&e^{i\gamma_3}
\end{array}
\right],
\label{ph7756}
\end{eqnarray}

and extract it directly from the mass matrix through the transformation:

\begin{eqnarray}
PMP^*=
\left[
\begin{array}{ccc}
a&xe^{i(\gamma_1-\gamma_2)}&ye^{i(\gamma_1-\gamma_3)}\\
x^*e^{-i(\gamma_1-\gamma_2)}&b&ze^{i(\gamma_2-\gamma_3)}\\
y^*e^{-i(\gamma_1-\gamma_3)}&z^*e^{-i(\gamma_2-\gamma_3)}&c
\end{array}
\right].
\label{some54545}
\end{eqnarray}

We denote:
\begin{eqnarray}
&&\gamma_1-\gamma_2=\eta+\tau
\nonumber\\
&&\gamma_1-\gamma_3=\eta-\tau
\nonumber\\
&&\gamma_2-\gamma_3=-2\tau
\label{den6665}
\end{eqnarray}

and further,
\begin{eqnarray}
&&\alpha_1=\beta_1+\tau
\nonumber\\
&&\alpha_2=\beta_2-\tau
\nonumber\\
&&\alpha_3=\beta_3-2\tau.
\label{deb6676}
\end{eqnarray}

This leads to the new phases for the elements x, y, z:
\begin{eqnarray}
&&|x|e^{i\beta_1}\rightarrow |x|e^{i(\alpha_1+\eta)}=x'
\nonumber\\
&&|y|e^{i\beta_2} \rightarrow |y|e^{i(\alpha_2+\eta)}=y'
\nonumber\\
&&|z|e^{i\beta_3} \rightarrow |z|e^{i\alpha_3}=z'.
\label{rel664554}
\end{eqnarray}

 Note that the phases $\eta$ and $\tau$ are introduced solely with the purpose of diagonalizing the mass matrix M with the "standard" form of the leptonic  matrix given in Eq. (\ref{lept77565}). We will work with the phases $\alpha_1$, $\alpha_2$, $\alpha_3$ and determine the extra phases $\eta$ and  $\tau$ from consistency conditions.

Now assume that the new mass matrix,
\begin{eqnarray}
M'=
\left[
\begin{array}{ccc}
a&x'&y'\\
(x')^*&b&z'\\
(y')^*&(z')^*&c
\end{array}
\right]
\label{new55443}
\end{eqnarray}

is diagonalized by the standard form of the leptonic mixing matrix. Here we make the simplifying assumption that the transformation matrix for the charged leptons is the unit matrix so that the phase P is absorbed in the charged lepton sector.
In case there is  a rotation matrix different from one then the matrix which realizes the diagonalization should be multiplied by the corresponding charged lepton matrix which should have the same form and  should be treated similarly.
We denote:
\begin{eqnarray}
U=
\left[
\begin{array}{ccc}
1&0&0\\
0&c_{23}&s_{23}\\
0&-s_{23}&c_{23}
\end{array}
\right]
\left[
\begin{array}{ccc}
c_{13}&0&s_{13}e^{-i\delta}\\
0&1&0\\
-s_{13}e^{i\delta}&0&c_{13}
\end{array}
\right]
\left[
\begin{array}{ccc}
c_{12}&s_{12}&0\\
-s_{12}&c_{12}&0\\
0&0&1
\end{array}
\right]=
R_{23}U_{13}R_{12}.
\label{lept77565}
\end{eqnarray}

Then the correct diagonalization procedure is,
\begin{eqnarray}
M'=UM_dU^{\dagger}
\label{dig66454}
\end{eqnarray}

which may be written also as,
\begin{eqnarray}
R_{23}^tM'R_{23}=U_{13}(R_{12}M_d R_{12}^T)U_{13}^{\dagger}.
\label{rel553443}
\end{eqnarray}

The eigenvalue matrix $M_d$ has the form:
\begin{eqnarray}
M_d=
\left[
\begin{array}{ccc}
m_1&0&0\\
0&m_2&0\\
0&0&m_3
\end{array}
\right].
\label{eig5546}
\end{eqnarray}

Here $m_i$ can be the masses or the square of the masses depending on the choice for the matrix M'.
Then from the elements (12) and (13) of Eq. (\ref{rel553443}) one obtains:
\begin{eqnarray}
&&\tan_{23}=\frac{x_2'}{y_2'}
\nonumber\\
&&\tan{\delta}=-\frac{x_2's_{23}+y_2'c_{23}}{x_1's_{23}+y_1'c_{23}}
\label{re4434}
\end{eqnarray}

where $x'=x_1'+ix_2'$, $y'=y_1'+iy_2'$ and $z'=z_1'+iz_2'$.
However Eq.(\ref{re4434}) is not sufficient by itself to determine the CP violating phase. We need a consistency condition to determine the  phase $\eta$. The
element (23) of Eq. (\ref{rel553443}) yields:
\begin{eqnarray}
\tan{\delta}=-\frac{z'_2}{c_{23}s_{23}(b-c)+(c_{23}^2-s_{23}^2)z_1'}
\label{t5554}
\end{eqnarray}

We can determine the phase $\eta$ from:
\begin{eqnarray}
-\frac{x_2's_{23}+y_2'c_{23}}{x_1's_{23}+y_1'c_{23}}=-\frac{z_2'}{c_{23}s_{23}(b-c)+(c_{23}^2-s_{23}^2)z_1'}
\label{eq332}
\end{eqnarray}

This is equivalent to:
\begin{eqnarray}
z_2'(x_1'x_2'+y_1'y_2')=y_2'x_2'(b-c)+(y_2^{'2}-x_2^{'2})z_1'
\label{eq221}
\end{eqnarray}

and leads to:
\begin{eqnarray}
&&-|x|^2|z|\cos(2\alpha_1+2\eta+\alpha_3)+|y|^2|z|\cos(2\alpha_2+2\eta-\alpha_3)+|x||y|(b-c)\cos(\alpha_1+\alpha_2+2\eta)=
\nonumber\\
&&|x||y|(b-c)\cos(\alpha_1-\alpha_2)-|z|\cos(\alpha_3)(|x|^2-|y|^2).
\label{tr4554}
\end{eqnarray}

We denote:
\begin{eqnarray}
&&u=-|x|^2|z|\cos(2\alpha_1+\alpha_3)+|y|^2|z|\cos(2\alpha_2-\alpha_3)+|x||y|(b-c)\cos(\alpha_1+\alpha_2)
\nonumber\\
&&v=|x|^2|z|\sin(2\alpha_1+\alpha_3)-|y|^2|z|\sin(2\alpha_2-\alpha_3)-|x||y|(b-c)\sin(\alpha_1+\alpha_2)
\nonumber\\
&&w=|x||y|(b-c)\cos(\alpha_1-\alpha_2)-|z|\cos(\alpha_3)(|x|^2-|y|^2)
\label{not665}
\end{eqnarray}

Then Eq.(\ref{tr4554}) can be written as:
\begin{eqnarray}
u\cos(2\eta)+v\sin(2\eta)=w
\label{eq33232}
\end{eqnarray}

with the solutions:
\begin{eqnarray}
&&\sin(2\eta)=\frac{vw\pm u\sqrt{u^2+v^2-w^2}}{u^2+v^2}
\nonumber\\
&&\cos(2\eta)=\frac{uw\mp v\sqrt{u^2+v^2-w^2}}{u^2+v^2}.
\label{not665}
\end{eqnarray}

Then one can extract the CP violation phase as:
\begin{eqnarray}
\tan(\delta)=
\frac{(|x|^2\cos(2\alpha_1)+|y|^2\cos(2\alpha_2))\frac{uw\mp v\sqrt{u^2+v^2-w^2}}{u^2+v^2}-(|x|^2\sin(2\alpha_1)+|y|^2\sin(2\alpha_2))\frac{vw\pm u\sqrt{u^2+v^2-w^2}}{u^2+v^2}-(|x|^2+|y|^2)}
{(|x|^2\sin(2\alpha_1)+|y|^2\sin(2\alpha_2))\frac{uw\mp v\sqrt{u^2+v^2-w^2}}{u^2+v^2}+(|x|^2\cos(2\alpha_1)+|y|^2\cos(2\alpha_2))\frac{vw\pm u\sqrt{u^2+v^2-w^2}}{u^2+v^2}}
\label{very7776}
\end{eqnarray}

Then from Eq. (\ref{re4434}) one can determine $\tan_{23}$:
\begin{eqnarray}
\tan_{23}=\pm\frac{|x|}{|y|}\sqrt{\frac{1-\cos(2\alpha_1)\cos(2\eta)+\sin(2\alpha_1)\sin(2\eta)}{1-\cos(2\alpha_2)\cos(2\eta)+\sin(2\alpha_2)\sin(2\eta)}}
\label{23444}
\end{eqnarray}

The ratio of the real parts of the elements (12) and (23) in Eq.(\ref{rel553443}) leads to $\tan_{13}$:
\begin{eqnarray}
\tan_{13}=\frac{|z|\sin(\alpha_3)(|x|^2+|y|^2-(|x|^2\cos(2\alpha_1)+|y|^2\cos(2\alpha_2))\cos(2\eta)+(|x|^2\sin(2\alpha_1)+|y|^2\sin(2\alpha_2))\sin(2\eta))^{1/2}}
{\sqrt{2}|x||y|\sin(\alpha_2-\alpha_1)\sin(\delta)}
\label{1344}
\end{eqnarray}

Furthermore the rest of the entries can be obtained  form the real part of the elements (22), (12), (13) in Eq. (\ref{rel553443}) and from the trace condition:

\begin{eqnarray}
&&m_3=(a+b+c)-m_1-m_2
\nonumber\\
&&s_{12}^2m_1+c_{12}^2m_2=c_{23}^2b+s_{23}^2c-2c_{23}s_{23}z_1'
\nonumber\\
&&c_{13}s_{12}c_{12}(m_2-m_1)=x_1'c_{23}-y_1's_{23}
\nonumber\\
&&x_1's_{23}+y_1'c_{23}=s_{13}c_{13}\cos(\delta)(a+b+c-m_1-m_2-c_{12}^2m_1-s_{12}^2m_2).
\label{rel775665}
\end{eqnarray}

By separating the angle $\theta_{12}$ and the masses $m_1$ and $m_2$ from the other parameters already determined we obtain the new system:
\begin{eqnarray}
&&s_{12}^2(m_1-m_2)+m_2=p
\nonumber\\
&&s_{12}^2(m_2-m_1)+m_2+2m_1=r
\nonumber\\
&&s_{12}c_{12}(m_2-m_1)=s
\label{new553443}
\end{eqnarray}

where by $p$, $r$ and $s$ we denote,
\begin{eqnarray}
&&p=c_{23}^2b+s_{23}^2c-2c_{23}s_{23}z_1'
\nonumber\\
&&r=\frac{x_1's_{23}+y_1'c_{23}}{s_{13}c_{13}\cos(\delta)}+(a+b+c)
\nonumber\\
&&s=\frac{x_1'c_{23}-y_1's_{23}}{c_{13}}.
\label{res442332}
\end{eqnarray}

The system in Eq. (\ref{res442332}) can be solved and leads to:
\begin{eqnarray}
&&m_1=\frac{1}{2}[\frac{r+p}{2}\pm\sqrt{(\frac{r}{2}-\frac{p}{2})^2+2(p^2+2s^2)}]
\nonumber\\
&&m_2=\frac{1}{2}[\frac{r+p}{2}\mp\sqrt{(\frac{r}{2}-\frac{p}{2})^2+2(p^2+2s^2)}]
\nonumber\\
&&\sin(2\theta_{12})=2s[\mp\sqrt{(\frac{r}{2}-\frac{p}{2})^2+2(p^2+2s^2)}]^{-1}
\label{res442332}
\end{eqnarray}

From a combination of the elements (11), (33), (13), (12) of the Eq. (\ref{rel553443}) we determine a final consistency condition:
\begin{eqnarray}
a-s_{23}^2b-c_{23}^2c-2s_{23}c_{23}z_1'=(1-\tan_{13}^2)\frac{x_1'y_2'-y_1'x_2'}{z_2'}.
\label{cos665775}
\end{eqnarray}

Eq. (\ref{cos665775}) is very complicated and cannot be solved exactly. However in the case of neutrinos one can make the  following simplifying assumptions \cite{Jora3}, \cite{Jora4}:

\begin{eqnarray}
&&|z|{\rm \,is\,of \,order\,} \rho
\nonumber\\
&&|y| {\rm\, is\,of\,order\,}\rho^2
\nonumber\\
&&|x|{\rm \,is\, of \,order\,} \rho^2
\nonumber\\
&&s_{13} {\rm \,is\,of\, order\,} \rho
\label{approx5546}
\end{eqnarray}

where $\rho$ is a small parameter. In order to have  a very good agreement with the experiment it is enough to consider angles only of order $\rho$ while the other parameters are considered of order $\rho^2$. Then from the last line in Eq. (\ref{approx5546}) and Eq. (\ref{1344}) one obtains:

\begin{eqnarray}
\sin{\alpha_3}{\rm \,is\,of\, order\,} \rho^2 \approx 0
\label{rel66454}
\end{eqnarray}

Note that in this approximation the presence of a nonzero CP violation phase  requires that the denominator in the r. h. s. of Eq. (\ref{t5554}) be zero which can be achieved by an appropriate choice of parameters.

Then condition (\ref{cos665775}) is  automatically satisfied,
\begin{eqnarray}
[a-s_{23}^2b-c_{23}^2c-2s_{23}c_{23}z_1']z_2\approx\rho^3\approx x_1'y_2-y_1'x_2.
\label{res4443}
\end{eqnarray}

 and all the angles are determined from:

\begin{eqnarray}
&&\tau=\frac{\beta_3}{2}
\nonumber\\
&&\alpha_1=\beta_1+\frac{\beta_3}{2}
\nonumber\\
&&\alpha_2=\beta_1-\frac{\beta_3}{2}.
\label{ew332}
\end{eqnarray}

This concludes the calculation of the masses and angles in terms of the initial parameters in the mass matrix.

An interesting analytical treatment of the neutrino masses and mixings in a different context has been given recently in \cite{King}.

\section{CP violation}

From Eq. (\ref{very7776}) and the calculations that lead  to it we determine that in order to have a CP phase zero the following conditions should be fulfilled:
\begin{eqnarray}
&&\alpha_3=k_1\pi
\nonumber\\
&&\alpha_1+\eta=k_2\pi
\nonumber\\
&&\alpha_2+\eta=k_3\pi,
\label{cond6645}
\end{eqnarray}

with $k_1$, $k_2$, $k_3$ integers.

Going back to Eq.(\ref{some54545}) we obtain that  the necessary and sufficient condition for a zero CP violating phase is:

\begin{eqnarray}
\beta_1-\beta_2+\beta_3=k\pi,
\label{cond554646}
\end{eqnarray}

where k is an integer.

Our result agrees to that obtained in \cite{Schechter1},\cite{Schechter2} from a different perspective.

We derive analytical formulas for the mixing angles and mass eigenvalues for the case when the leptonic mixing matrix is given simply  by the neutrino mixing matrix.
These formulas may prove useful in constructing and analyzing various models for neutrino masses, mixings or CP violation.

\section*{Acknowledgments} \vskip -.5cm

The work of R. J. was supported by PN 09370102/2009. The work of J. S. was supported in part by the US DOE under Contract No. DE-FG-02-85ER 40231.

\end{document}